\documentclass[12pt]{article}
\usepackage{graphics}
\usepackage{amssymb}

\newcommand{\rf}[1]{(\ref{#1})}
\newcommand{\beq}{\begin{equation}}
\newcommand{\eeq}{\end{equation}}
\newcommand{\bea}{\begin{eqnarray}}
\newcommand{\eea}{\end{eqnarray}}

\newcommand{\e}{\mbox{e}}
\renewcommand{\d}{\mbox{d}}
\newcommand{\del}{\delta}

\newcommand{\oh}{\frac{1}{2}}

\newcommand{\tr}{\mathrm{tr}\,}
\newcommand{\ra}{\rangle}
\newcommand{\la}{\langle}

\newcommand{\mi}{\!-\!}
\newcommand{\equ}{\!=\!}
\newcommand{\pl}{\!+\!}

\newcommand{\cT}{{\cal T}}

\newcommand{\cN}{{\cal N}}

\newcommand{\cO}{{\cal O}}

\newcommand{\hT}{{\hat{T}}}
\newcommand{\hH}{{\hat{H}}}

\begin{document}

{\normalsize \hfill SPIN-2003/25}\\
\vspace{-1.7cm}
{\normalsize \hfill ITP-UU-03/39}\\

\begin{center}
\vspace{3.5cm}
{\Large
\bf Renormalization of 3d quantum gravity\\
\vspace{0.4cm}
from matrix models}

\vspace{2.0cm}

{\large\sl J. Ambj\o rn}$\,^{a}$,
{\large\sl J. Jurkiewicz}$\,^{b}$
{\large and} {\large\sl R. Loll}$\,^{c}$  

\vspace{1.2cm}
{\footnotesize

$^a$~The Niels Bohr Institute, Copenhagen University\\
Blegdamsvej 17, DK-2100 Copenhagen \O , Denmark.\\
{\tt email: ambjorn@nbi.dk}\\

\vspace{10pt}

$^b$~Institute of Physics, Jagellonian University,\\
Reymonta 4, PL 30-059 Krakow, Poland.\\
{\tt email: jurkiewi@thrisc.if.uj.edu.pl}\\

\vspace{10pt}

$^c$~Spinoza Institute and Institute for Theoretical Physics, Utrecht University, \\
Leuvenlaan 4, NL-3584 CE Utrecht, The Netherlands.\\ 
{\tt email:  loll@phys.uu.nl}\\

\vspace{10pt}
}
\vspace{48pt}

\end{center}

\begin{center}
{\large\bf Abstract}
\end{center}

Lorentzian simplicial quantum gravity is a non-perturbatively
defined theory of quantum gravity which predicts a positive cosmological
constant. Since the approach is based
on a sum over space-time histories, it is perturbatively
non-renormalizable even in three dimensions. By mapping the three-dimensional
theory to a two-matrix model with ABAB interaction we show that
both the cosmological and the (perturbatively) non-renormalizable gravitational  
coupling constant undergo additive renormalizations consistent with 
canonical quantization.

\vspace{12pt}
\noindent


\newpage

\subsection*{Introduction}\label{intro}
\vspace{.4cm}
Defining a theory of quantum gravity as a suitable sum over space-time 
histories is an appealing proposition, since it can in principle be done in a completely
background-independent and non-perturbative way, with the structure of space-time
being determined {\it dynamically}. In two space-time dimensions, such a program
can be carried out successfully, although in this case -- because of the
absence of propagating gravitons -- it may be more appropriate to 
talk about a theory of ``quantum geometry" rather than one of quantum gravity. 
A well-known example is the non-perturbative lattice formulation 
of 2d (Euclidean) gravity which reproduces quantum Liouville theory in the limit of 
vanishing lattice spacing \cite{david1,david2,distler}. 
Attempts to use similar combinatorial and matrix-model 
techniques to extract information about the non-perturbative structure 
of higher-dimensional gravity have until recently met with little success. 
However, {\it if one performs the sum over geometries over
space-times of Lorentzian (as opposed to Riemannian) signature}, matrix-model 
methods {\it can} be applied profitably in the non-perturbative 
quantization of three-dimensional quantum gravity, as was first shown in \cite{ajlv}.
This line of investigation will be pursued further in the present work.

Quantum gravity in three space-time dimensions represents an interesting case
in between dimensions two and four. On the one hand, it contains 
no propagating gravitational degrees of freedom and can be reduced
classically to a finite-dimensional physical
phase space, both in a metric \cite{moncrief} and a connection (Chern-Simons)
formulation \cite{witten}.\footnote{Whether and to what extent the associated 
{\it quantum} theories are related is still a contentious issue.}    
Nevertheless, the unreduced theory in terms of the metric $g_{\mu\nu}$ 
appears to be non-renormalizable
when one tries to expand around a fixed background geometry, just as
in four dimensions. A definition of three-dimensional quantum gravity 
via a  ``sum over geometries'' therefore seems to require a genuinely
non-perturbative construction, and in turn may shed 
light on the problem of non-renormalizability of the full,
four-dimensional theory, where an explicit classical reduction 
is not available.

A non-perturbative definition of the sum over geometries in 
three- and four-dimensional quantum gravity was proposed 
in \cite{ajl1,d3d4}. Unlike previous approaches,
this method of ``Lorentzian dynamical triangulations"
or ``Lorentzian simplicial quantum gravity" 
uses space-time geometries with physical, Lorentzian
signature, rather than positive-definite Riemannian geometries
as a fundamental input. Details on the classes of geometries included in
the path sum and on earlier two-dimensional work that provided the motivation
for this approach can be found in \cite{al,d3d4,badhonnef}. 
In view of the recent observational progress in cosmology (see \cite{cosmol} 
for a recent review)
we should point out that the physical, renormalized cosmological constant 
in all of these models is necessarily positive. 

In this paper, we will present an explicit analysis of the renormalization
behaviour of the 3d Lorentzian model, using a matrix-model
formulation. This follows previous work which analyzed the phase
structure of three-dimensional quantum gravity (for spherical spatial
topology) with the help of
computer simulations \cite{ajl2,ajl3,ajl4}, and a demonstration \cite{ajlv} 
that 3d Lorentzian
dynamical triangulations can be mapped to graph configurations
generated by the so-called ABAB-matrix model \cite{kz}.

Within {\it continuum} approaches to quantum gravity there have also
been attempts to prove the non-perturbative renormalizability of gravity
beyond dimension two, starting with an analysis of the theory
in $2\pl\epsilon$ dimensions \cite{kawai1,kawai2,kawai3}. More
recently, an effective average action approach has produced
evidence of a non-trivial fixed point through an analysis of
renormalization group flow equations \cite{reuter1,reuter2,reuter3}.
\vspace{.4cm}

\subsection*{Quantum gravity and the ABAB-matrix model}\label{model}
\vspace{.4cm}
We start out with a brief description of
the three-dimensional Lorentzian simplicial space-times
appearing in the sum over geometries, and the construction of the
partition function. In the standard formulation of the model, the spatial 
hypersurfaces of constant integer proper time $t$ are given by
two-dimensional equilateral triangulations, each corresponding to a unique
piecewise flat 2d geometry. These are the same geometries as appear in 
the construction of 2d Euclidean quantum gravity, which is known to be
rather robust with regard to changes in both the types of building blocks
used and their gluing rules \cite{robust}. We exploited this universality in
\cite{ajlv} by using 2d spatial geometries made up of equilateral squares
instead of triangles,
and accordingly changing the 3d building blocks from tetrahedra only
to a set of tetrahedra and pyramids.

\begin{figure}[t]
\centerline{\scalebox{0.6}{\rotatebox{0}
{\includegraphics{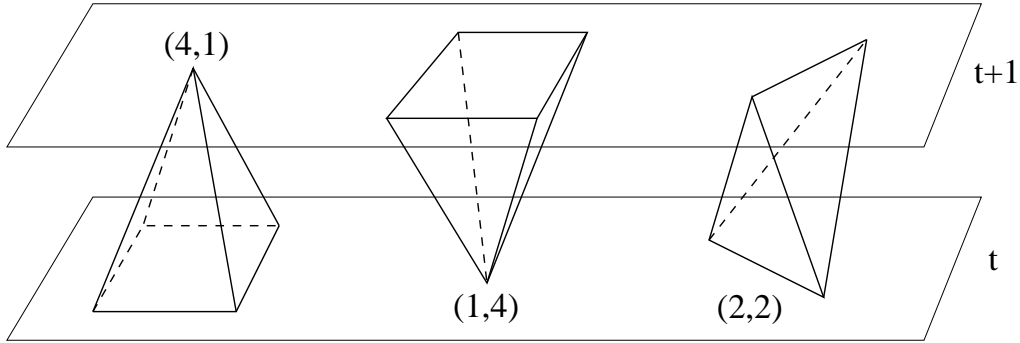}}}}
\caption[pyramids]{The fundamental building blocks of 3d Lorentzian quantum
gravity interpolate between adjacent spatial slices of integer times $t$ and
$t\pl 1$, and are
labelled according to the numbers $(i_t,i_{t\pl 1})$ of their vertices lying in the two slices.}
\label{pyramids}
\end{figure}
Any two neighbouring spatial quadrangulations at times $t$ and $t\pl 1$ can
be connected (in many inequivalent ways) by a three-dimensional 
``sandwich" geometry constructed from these building blocks, as indicated in
Fig.\ref{pyramids}. The square base of a pyramid (or an upside-down pyramid) 
coincides with a square of the spatial slice at time $t$ (or $t\pl 1$), whereas
the tetrahedral building block is needed to connect between the two types of
pyramids within the same sandwich. 

The amplitude for propagation from an initial quadrangulation $g_1$ to a final
one $g_2$ in $n$ proper-time steps is obtained by summing over all geometrically
distinct ways of stacking $n$ sandwich geometries $\Delta t\equ 1$ in between
$g_1$ and $g_2$,
in such a way that their 2d boundary geometries match pairwise at integer times.
The weight of each geometry is given by a discretized version of the Einstein action, 
here conveniently taken as the Regge action for piecewise linear geometries \cite{regge}. 
After Wick-rotating, the partition function (or proper-time propagator) can be written as 
\beq\label{lq.1}
Z(\kappa,\lambda;g_1,g_2,n)= \sum_{\cT, \partial \cT = {g_1\cup g_2}} 
\frac{1}{C_\cT}\; \e^{-S(\cT)},
\eeq
where $C_\cT$ is the order of the automorphism group of the (generalized) 
triangulation $\cT$, 
and the sum is over all $\cT$ with fixed boundaries $g_1$ and $g_2$
of the kind just described.
The gravitational action, including a cosmological term, is given by
\beq\label{lq.2}
S(\cT) = -\kappa \Big(N_{14}(\cT)\pl N_{41}(\cT)\mi N_{22}(\cT)\Big)
\pl\lambda \Big(N_{14}(\cT)\pl N_{41}(\cT)\pl \oh N_{22}(\cT)\Big),
\eeq
where $N_{41}(\cT)$ and $N_{14}(\cT)$ count the numbers of pyramids 
and upside-down pyramids and $N_{22}(\cT)$ the number of tetrahedra contained 
in a given triangulation $\cT$. The simplicity of the 
Regge action in our case stems from the fact that we use only two types of 
building blocks, and contributions to volumes and curvatures (in the
form of deficit angles) occur only in terms of a few basic units
(see \cite{d3d4,ajlv} for further details). The simplicial action contains two 
dimensionless coupling constants $\kappa$ and $\lambda$, related 
to their continuum counterparts by\footnote{Note that our cosmological constant 
$\Lambda^{(0)}$ is defined as the quantity that multiplies the volume term 
$\int \d ^3 x \, \sqrt{g}$. More conventionally this term would be
called $ \Lambda^{(0)}/(8\pi G^{(0)})$.}  
\beq\label{II.3}
\kappa= \frac{a}{4\pi G^{(0)}}\Big( -\pi+3\cos^{-1}\frac{1}{3}  \Big),~~~
{\lambda} = \frac{a^3 \Lambda^{(0)}}{24\sqrt{2}\pi},
\eeq 
where $a$ is a geodesic lattice cut-off with the dimension of length.
It should be emphasized that these are ``na\"{\i}ve'' relations
between the dimensionless lattice coupling constants and those of
the continuum theory, which will {\it not} be valid in the quantum theory. 
As we shall see in due course, additive renormalizations
of both coupling constants will be needed in that case.

\begin{figure}[t]
\centerline{\scalebox{0.7}{\rotatebox{0}{\includegraphics{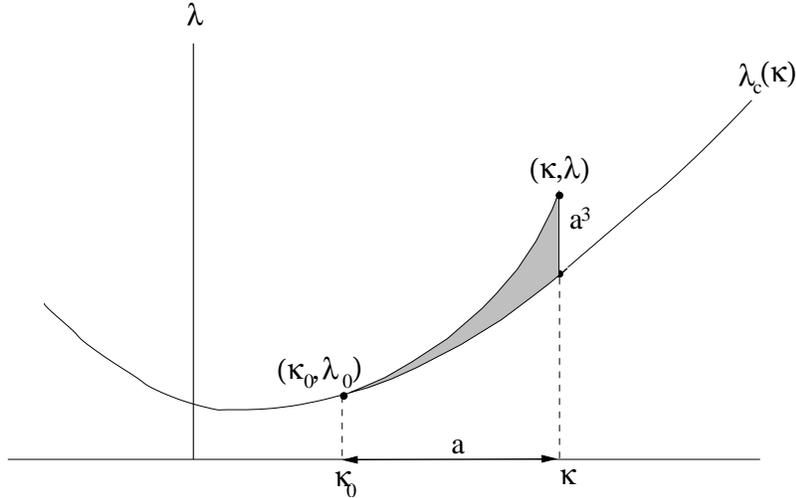}}}}
\caption[renormkl]{The phase diagram of 3d Lorentzian quantum gravity in the
plane spanned by the bare inverse gravitational coupling $\kappa$ and
the bare cosmological constant $\lambda$, together with the canonical 
approach to a point $(\kappa_0,\lambda_c(\kappa_0))$ on the critical line.}
\label{renormkl}
\end{figure}

We can rewrite the partition function \rf{lq.1} as
\beq\label{r.1}
Z(\kappa,\lambda;g_1,g_2,n) = 
\sum_N \e^{-\lambda N} \sum_{\cT_N} \frac{1}{C_{\cT_N}}\;
\e^{ \kappa (N_{14}(\cT_N)+N_{41}(\cT_N)-N_{22}(\cT_N))},
\eeq
where the sum over the total space-time volume
$N = N_{14}+N_{41}+\oh N_{22}$ has been pulled out, together with
the accompanying Boltzmann weight $\e^{-\lambda N}$, and the remaining sum 
runs over all triangulations $\cT_N$ of fixed volume $N$, whose
Boltzmann weights depend on the curvature term multiplying $\kappa$.
To leading order, the number of triangulations at fixed volume grows
exponentially with the volume, leading to the asymptotic behaviour
\beq\label{r.2}
f(N; g_1,g_2) \, \e^{\lambda_c(\kappa) N},
\eeq
for the second sum in (\ref{r.1}), where $f(N;g_1,g_2)$ indicates subleading terms 
in $N$. It follows immediately that for a given $\kappa$ the regularized quantum gravity 
model is only well defined (that is, its state sum converges) for $\lambda > \lambda_c(\kappa)$,
corresponding to the region above the critical line in the phase diagram of
Fig. \ref{renormkl}. 
The critical line limits the region of convergence of the partition function $Z$. Taking
$\lambda\to\lambda_c(\kappa)$ from {\it inside} this region of convergence,
the average value of (suitable powers of) $N$
will diverge, corresponding to the limit of infinite lattice volume. 
Such a limit is clearly necessary if a continuum 
limit in any conventional sense is to be achieved.

The continuum limit is obtained 
by scaling the lattice spacing $a$ to zero while keeping the continuum time 
$T \equ n\cdot a$ fixed (and therefore, increasing the number $n$ of discrete
time steps at a rate $1/a$). Different, non-canonical scaling relations between 
$T$ and $a$ are in principle possible\footnote{In two-dimensional 
{\it Euclidean} quantum gravity the
proper time $T$ scales anomalously and one has to keep $n\sqrt{a}$ fixed \cite{scaling}.
By contrast, the scaling in two-dimensional Lorentzian simplicial quantum 
gravity is canonical \cite{al}. The relation between the 
two formulations is well understood \cite{ackl}.},
but the computer simulations of \cite{ajl2} 
supported the presence of canonical scaling in 3d quantum gravity. 
More precisely, we expect to leading order in $a$ a scaling of the form
\beq\label{lq.3}
\frac{a}{G} = \kappa -\kappa_0,~~~~a^3 \Lambda = \lambda(\kappa)-\lambda_c(\kappa),
\eeq
as illustrated in Fig. \ref{renormkl}. The approach to the critical line is 
governed by the dimensionless combination $G^3\Lambda$ which serves as the 
true, ``observable'' coupling constant of 3d quantum gravity.
The physics underlying \rf{lq.3} is as follows: for a given 
value of the bare inverse gravitational coupling 
$\kappa$ the average discrete space-time volume $\la N \ra$ and its
dimensionful counterpart $\la V \ra$ behave like 
\beq\label{lq.4}
\la N \ra \sim \frac{1}{\lambda-\lambda_c(\kappa)}~~~~\Longrightarrow ~~~~~ 
\la V \ra := a^3 \la N \ra \sim \frac{a^3}{\lambda -\lambda_c(\kappa)},
\eeq
that is, the number of building blocks diverges in the limit as 
$\lambda \to \lambda_c(\kappa)$. 
The physical requirement that the continuum volume $\la V \ra$ remain finite 
and be proportional to the inverse {\it renormalized} cosmological constant
$1/\Lambda$ fixes the second scaling relation in \rf{lq.3}.
The first relation is then determined by demanding that $G^3 \Lambda$ be a dimensionless
coupling constant of the theory. This is precisely achieved by approaching
a given point $(\kappa_0,\lambda_c(\kappa_0))$ on the critical curve according to the 
canonical scaling assignment \rf{lq.3}. Note in passing that there is no
way of obtaining a renormalized cosmological coupling $\Lambda\leq 0$, in
agreement with our earlier remarks. Also, we choose the approach
to the critical line such that the sign of the renormalized Newton constant 
is standard and positive.

Our construction raises the question of whether or not
physics depends on the choice of $\kappa_0$. Indications from
the computer simulations of the model are that the final result 
is independent of the value of $\kappa_0$ in the range probed \cite{ajl2}.
We will discuss in the following how this question can be addressed  analytically. 

Let $g_t$ and $g_{t+1}$ be two spatial quadrangulations at $t$ and $t\pl 1$,
and $\la g_{t+1}|\hT |g_t\ra$ the 
transition amplitude or proper-time propagator for the single time step from $t$ to 
$t\pl 1$. By definition, $\hT$ is 
the transfer matrix in the sense of Euclidean lattice theory, and can be shown
to satisfy the usual properties of a transfer matrix \cite{d3d4}. The 
propagator for $n$ time steps is obtained by an $n$-fold iteration,
\beq\label{lq.5}
Z(\kappa,\lambda;g_1,g_2,n) = \la g_2| \hT^n |g_1\ra.
\eeq                                       

Consider now the matrix model of two hermitian $M\times M$-matrices with
partition function
\beq\label{II.0}
Z(\alpha_1,\alpha_2,\beta) = \int \d A \d B \; \e^{-M \tr (A^2+B^2 -
\frac{\alpha_1}{4} A^4-\frac{\alpha_2}{4} B^4 -\frac{\beta}{2} ABAB)}.
\eeq
In the context of the large-$M$ expansion the free energy $F$ can be
expressed as
\beq\label{II.0a}
M^2 F(\alpha_1,\alpha_2,\beta) \equiv - \log Z(\alpha_1,\alpha_2,\beta) = 
\sum_{h=0}^{\infty} M^{\chi(h)} F_h(\alpha_1,\alpha_2,\beta),
\eeq 
where $\chi(h) = 2-2h$ is the Euler number of the quadrangulations
dual to the four-valent graphs generated by the matrix model. 
It was argued in \cite{ajlv} that  the transfer matrix for transitions between 
two spatial geometries $g_t$ and $g_{t+1}$ of genus $h$
is related to $F_h(\alpha_1,\alpha_2,\beta)$ according to
\beq\label{II.1}
F_h(\alpha_1,\alpha_2,\beta) = \sum_{N_t,N_{t+1}} \e^{-z_t N_t-z_{t+1}N_{t+1}} 
\!\!\!\!\!
\sum_{g_{t+1}(N_{t+1}),g_{t}(N_t)} \la g_{t+1}(N_{t+1})| 
\hT |g_t (N_t)\ra_h,
\eeq
where $N_t$ and $N_{t+1}$ denote the numbers of squares of the 
quadrangulations defining the spatial 
geometries at times $t$ and $t+1$,  both of Euler number $\chi(h)$. 
Pulling out the double-sum over discrete boundary volumes is convenient
when studying the transfer matrix {\it per se} (see \cite{al,iceland} for an
analogous procedure in two space-time dimensions). 
The two dimensionless boundary constants $z_t$ and $z_{t+1}$ can be 
viewed as  cosmological coupling constants for the boundary areas.
For the purposes of the present paper we will choose particular values 
for $z_t$ and $z_{t+1}$, in such a way that the relations
\beq\label{II.2}
\alpha_1 = \alpha_2 = \e^{\kappa-\lambda},~~~~~~~~~
\beta= \e^{-(\oh \lambda+\kappa)},
\eeq
hold between the matrix model coupling constants $\alpha_i$, $\beta$, 
and the bare gravitational and cosmological coupling constants $1/\kappa$ and $\lambda$ of
three-dimensional gravity. The relations (\ref{II.2}) were derived previously
in \cite{ajlv}, and we will use them in the next section to translate the
canonical approach \rf{lq.3} to the matrix model and draw conclusions
about the renormalization behaviour of the theory.

The derivation of eq.\ \rf{II.2} requires some explanation. Generic matrix 
elements of $\hat T$ in \rf{II.1} grow exponentially with the total discrete 
three-volume $N= N_t\pl N_{t+1}\pl N_{22}/2$, 
reflecting the fact that there are exponentially many three-geometries which 
interpolate between two given two-geometries $g_t$ and $g_{t+1}$.
This exponential growth is taken care of by the combined {\it additive renormalizations}
of the cosmological and gravitational constants, as discussed  earlier in this section. 

There is a completely analogous entropy for the boundary two-geometries,
since the number of quadrangulations of a given 
topology and a given discrete two-volume $N_t$ grows exponentially
with $N_t$. Just as in the case of the three-volume, this 
exponential growth can be cancelled by an {\it additive renormalization},
in this case
of the boundary cosmological constant $z_t$, leading to a renormalized
boundary cosmological constant multiplying a continuum area.
Assume that the second sum in \rf{II.1}
grows like $e^{z_c(N_t+N_{t+1})}$ to leading order in the boundary 
two-volumes, and renormalize $z_t$ and $z_{t+1}$ canonically
according to 
\beq\label{bcc}
z_t = z_c+a^2 Z_t,~~~~z_{t+1} = z_c+a^2 Z_{t+1}.
\eeq
Defining the continuum area $A_t$ of a quadrangulation of $N_t$ squares 
by $A_t:= N_t a^2$, the total area contribution in the exponential 
in \rf{II.1} becomes
\beq\label{bcc1}
(z_c-z_t)N_t+(z_c-z_{t+1})N_{t+1} = -(Z_t A_t+Z_{t+1}A_{t+1}),
\eeq
as anticipated. In this article, we set $Z_t=Z_{t+1}=0$, corresponding to 
$z_t=z_{t+1}=z_c$ in \rf{II.1}, since we are only
interested in the bulk coupling constants $\Lambda$ and $G$. This implies
the symmetry $\alpha_1=\alpha_2$, as well as the relation \rf{II.2}. From a technical 
point of view it means that we have to deal only with 
the symmetric ABAB-matrix model which, contrary to the asymmetric 
model, has been solved explicitly \cite{kz}.
\vspace{.4cm}

\subsection*{Renormalization of 3d gravity}\label{renormalization}
\begin{figure}[t]
\centerline{\scalebox{0.7}{\rotatebox{0}{\includegraphics{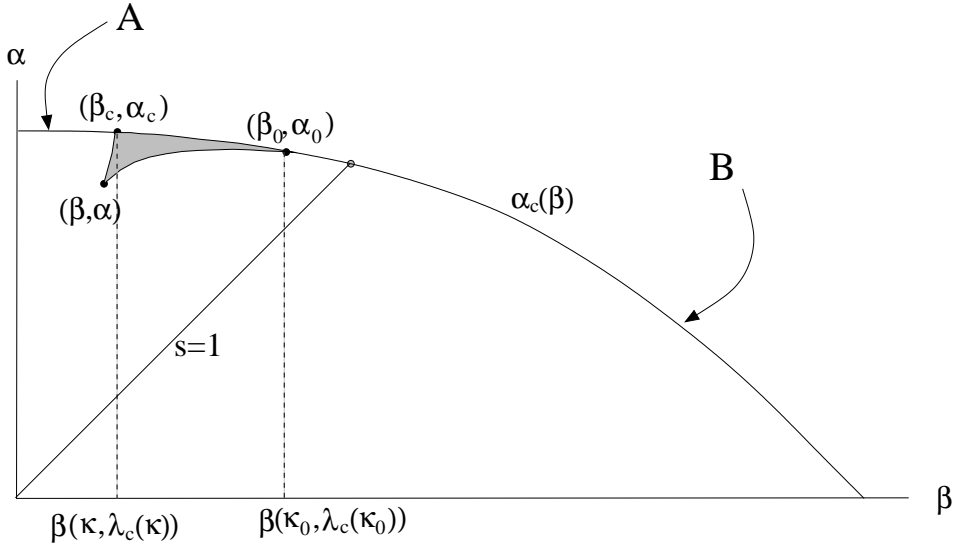}}}}
\caption[renormab]{The phase diagram of 3d Lorentzian quantum gravity in the
plane spanned by the two coupling constants $\beta$ and $\alpha$ of the matrix
model, together with the canonical 
approach to a point $(\beta_0,\alpha_0)$ on the critical line.
The end point $(\beta_c,\alpha_c=\beta_c)$ of the diagonal $s\equ 1$ 
separates phase A from phase B.}
\label{renormab}
\end{figure}

\vspace{.4cm}
The canonical approach \rf{lq.3} to a critical point 
$(\kappa_0,\lambda_0)$ on the critical line of the $(\kappa,\lambda)$-coupling
constant plane, Fig.\ref{renormkl}, can be mapped via \rf{II.2} to the 
$(\beta,\alpha)$-plane, as shown in Fig.\ref{renormab}. 
Let $F(\alpha,\beta)$ denote the free energy of the symmetric ABAB-matrix model, 
and set $\alpha_1=\alpha_2\equiv \alpha$. It is convenient to change variables from 
$(\beta,\alpha)$ to $(s,r)$, where
\beq\label{rs}
s = \frac{\beta}{\alpha},~~~~~r= \sqrt{\alpha^2+\beta^2}.
\eeq 
The upper right-hand quadrant of the $\alpha$-$\beta$-plane corresponds to
$r,s\in [0,\infty ]$. Approaching a point $(\beta_c(s),\alpha_c(s))$ on the critical 
line from below 
along a line segment of constant $s$, the coordinate $r$ will vary
between 0 and $r_c(s)\equ \sqrt{\alpha_c(s)^2+\beta_c(s)^2}$.
According to \cite{kz}, $F(\alpha,\beta)$ or $F(s,r)$ are analytic 
functions of their arguments {\it below} the critical line. Moreover,
approaching the critical line along $s\equ const$, $F(r,s)$ has 
an expansion
\beq\label{V.3}
F(s,r) -F(s,r_c(s)) = c_1(s) \del r + c_2(s) \del r^2 + c_{5/2}(s) \del r^{5/2}
+c_3(s) \del r^3 +\cdots
\eeq  
in the vicinity of the critical point $(s,r_c(s))$,
where $\del r \equ r_c(s)\mi r$ and where the coefficients $c_i(s)$ are
analytic functions of $s$ for both $0<s<1$ and $1< s <\infty$.
Around the special point $(s,r_c(s)) \equ (1, r_c(1))$ which separates the 
so-called A-phase ($s < 1$) from the B-phase ($s >1$), the behaviour is more 
complicated than the one given in \rf{V.3}. 
As discussed in \cite{ajlv}, phase A is the one 
relevant for canonical quantum gravity and we will consider only
coupling constant variations inside phase A.

The straight approach along $s=const$ to the critical line underlying 
\rf{V.3} is not the one relevant for three-dimensional
quantum gravity, since it would translate to a curve 
in the $(\kappa,\lambda)$-plane which approaches 
the corresponding critical point $(\kappa_0,\lambda_0)$ non-tangentially.
In the notation of \rf{lq.3}, this would imply
$\kappa-\kappa_0 \propto \lambda(\kappa)-\lambda_c(\kappa)$, in contradiction with the scaling relations 
\rf{lq.3}. Stated differently, insisting on canonical
dimensions for $G$ and $\Lambda$ and a finite $\Lambda$, the gravitational coupling 
$G$ would have to go to infinity 
like $1/a^2$ when the cut-off is removed. 

One can of course repeat the analysis of \cite{kz} for an arbitrary 
approach to the critical line. However, rather than giving 
the technical details of this, let us just state the final result 
for the case at hand.
We can approach a critical point $(\beta_0,\alpha_0)$ along 
any curve $(\beta(a),\alpha(a))$, where for convenience we
have identified the curve parameter $a$ with the lattice cut-off. 
For the canonical gravitational interpretation to be valid, the
scaling must follow \rf{lq.3}, that is, both the 
tangent and the curvature of the curve $(\beta(a),\alpha(a))$ must agree
with those of the critical line $(\beta_c(s),\alpha_c(s))$ at the 
point $(\beta_0,\alpha_0)$. The difference between the two curves
will only appear in their third-order derivatives, as indicated
by Fig.\ref{renormab}. 
In order to investigate the analyticity properties of the free energy,
we perform a decomposition
\beq\label{V.4}
F(\alpha(a),\beta(a)) -F(\alpha_0,\beta_0)= \Big(F(\alpha,\beta)-F(\alpha_c,\beta_c)\Big)+ 
\Big(F(\alpha_c,\beta_c)-F(\alpha_0,\beta_0)\Big),
\eeq
where, in the notation of Fig.\ref{renormab}, the approaching curve
$(\kappa(a),\lambda(a))$ translates into $(\beta(a),\alpha(a))$,
$(\beta_c,\alpha_c)$ corresponds to the point $(\kappa,\lambda_c(\kappa))$,
and $(\beta_0,\alpha_0)$ to $(\kappa_0,\lambda_0)$ on the critical line.
To evaluate the first difference in \rf{V.4} we can use
\beq\label{V.5}
\alpha -\alpha_c \sim \Lambda a^3+\cdots,~~~~\beta-\beta_c \sim \Lambda a^3+\cdots,
\eeq
as well as the expansion \rf{V.3}. In the second difference
we can use 
\beq\label{V.6}
\alpha_c-\alpha_0 \sim - a/G +\cdots,~~~~\beta_c-\beta_0 \sim -a/G+ \cdots,
\eeq
{\it without} any reference to the renormalized cosmological constant $\Lambda$,
defined by \rf{lq.3}. This happens because both 
$(\beta_0,\alpha_0)$ and $(\beta_c,\alpha_c)$ lie on the critical 
line, whereas $\Lambda$ is a measure of the {\it distance from} 
the critical line. The important point is that -- as long as we stay in phase A --
the difference $F(\alpha_c,\beta_c)\mi F(\alpha_0,\beta_0)$ is entirely analytic in 
$\alpha_c\mi\alpha_0$. We conclude that the non-analytic behaviour of the
free energy occurs as a function of the cosmological 
coupling constant alone. This non-analyticity ensures the existence of
an infinite-volume limit of 3d quantum gravity in the sense of \rf{lq.4}. 
{\it The renormalized gravitational coupling constant $G$ plays no role in 
taking the continuum limit}, which is entirely
dictated by the non-analytic part of $F(\alpha,\beta)$. 

Let us discuss this behaviour in some more detail. The free energy 
$F(\alpha,\beta)$ of the matrix model serves as the {\it partition function} of the 
sum over sandwich configurations of the three-dimensional 
Lorentzian gravity model, as described above. Its continuum limit 
is associated with a limit where the number $N$ of 3d building blocks
diverges, and $a\to 0$, while keeping the continuum three-volume 
$V= Na^3$ finite. 
This large-$N$ behaviour is related to the expansion 
\beq\label{expand1}
F(\alpha,\beta) = \sum_{N_{14},N_{41},N_{22}} \cN(N_{14},N_{41};N_{22})
\;\alpha^{N_{14}+N_{41}}\beta^{N_{22}},
\eeq
of $F(\alpha,\beta)$ into large powers of $\alpha$ and $\beta$, where $\cN(N_{14},N_{41};N_{22})$ 
denotes the number of three-geometries constructed from $(N_{14},N_{41},N_{22})$ 
building  blocks between neighbouring spatial surfaces at $t$ and $t\pl 1$ 
(see \cite{ajlv} for details). The non-analytic part of $F(\alpha,\beta)$ is associated with
simultaneous infinitely large 
powers of $\alpha$ and $\beta$, which in turn is reflected in a finite radius of convergence
of the power expansion. 

We will denote the non-analytic part of $F(\alpha,\beta)$ by $F_{singular}(\alpha,\beta)$, 
and it is only this part that should be kept when discussing the continuum 
limit. Thus, returning to the expansion \rf{V.3}, the first two terms on the 
right-hand side are irrelevant to a potential continuum limit dictated by the 
non-analytic term $(r_c -r)^{5/2}$.  
Likewise, the term $F(\alpha_c,\beta_c)\mi F(\alpha_0,\beta_0)$ in eq.\ \rf{V.4} 
can be ignored when discussing continuum physics. 
The term $F(\alpha,\beta)\mi F(\alpha_c,\beta_c)$ in that relation
is similar to the quantity \rf{V.3} which characterizes the non-tangential 
approach to a critical point. The continuum expression which survives
is therefore
\beq\label{V.10}
F_{singular}(\Lambda,G) \sim \Big( \Lambda a^3 \Big)^{5/2}.
\eeq
One would obtain the same expression in the 2d (Euclidean) quantum 
gravity interpretation given in \cite{kz}, except that 
the power of the lattice cut-off would be different. This is due to 
the tangential approach to the critical point in the present
case, reflecting the different physical properties of the higher-dimensional 
gravity theory.   

One should keep in mind that $F_{singular}$ 
is not identical with the partition function \rf{r.1} for three-dimensional
quantum gravity for $n\equ 1$, but rather is a particular sum of matrix 
elements of the transfer matrix between two adjacent constant proper-time
slices, which are separated by one lattice unit $a$. However, as was also 
argued in \cite{ajlv}, the study  
of this sum is sufficient to exhibit the renormalization behaviour
of the bare gravitational and cosmological coupling constants.\footnote{In
an analogous analysis of two-dimensional simplicial 
Lorentzian quantum gravity one also can deduce the renormalization
of the cosmological constant from the study of the same
restricted combination of matrix elements.}
The only way in which the (perturbatively)
non-renormalizable gravitational coupling constant $G$ makes an
appearance in 3d Lorentzian quantum gravity 
is by fixing the approach to the chosen critical point $\kappa_0$, and thereby
defining the dimensionless quantity
\beq\label{approach}
\frac{\lambda -\lambda_c(\kappa)}{(\kappa-\kappa_0)^3}= const. = \Lambda G^3.
\eeq
Consequently, all observables we may think of calculating in this
formulation will be of the form
\beq\label{approach1}
\cO (\Lambda,G) = \Lambda^{dim/3} F( \Lambda G^3)
\eeq
after the continuum limit has been performed, 
where {\it ``dim''} refers to the mass dimension of the observable $\cO$.
\vspace{.4cm}

\subsection*{Discussion}
\vspace{.4cm}
Three-dimensional simplicial Lorentzian quantum gravity gives an
explicit realization of the summation over three-geometries.
As in all quantum theories with a cut-off, a prescription must be given
of how to remove the cut-off and recover the underlying continuum
quantum field theory; we did this by specifying the renormalization of 
the bare coupling constants of the theory. The relation of the 
model to the ABAB-matrix model allowed 
us to give a detailed discussion of a possible renormalization
of the gravitational and cosmological coupling constants, 
consistent both with 
the existence of an infinite-volume limit of the model and 
with a canonical scaling of the renormalized coupling constants.

The bare gravitational and the bare 
cosmological coupling constants turned out to be subject to additive 
renormalizations. The perturbative non-renormalizability of the 
gravitational coupling constant is resolved in this 
non-perturbative approach by the fact that the {\it renormalized} 
gravitational coupling constant only appears in the particular
combination \rf{approach}, defined by the canonical approach to the 
critical line. 

One way to obtain more detailed information about the continuum limit 
would be by analyzing the full transfer matrix, instead of the contracted
version we have studied in the present work. From the transfer matrix
one can extract the {\it continuum} proper-time Hamiltonian $\hat H$ 
by virtue of the relation
\beq\label{tranf}
\hT = \e^{-a \hH} \approx \hat{I}-a \hH.
\eeq
This can be done explicitly in both two-dimensional Lorentzian and 
Euclidean simplicial quantum gravity, where the Hamiltonian is
a differential operator in a single variable, the one-volume of the
spatial universe.
Three-dimensional quantum gravity is more involved since
the spatial geometries at a fixed time
constitute an infinite-dimensional field space, 
spanned by the conformal factor and a finite number of 
Teichm\"{u}ller parameters. However, from our knowledge of the classical, 
canonical structure of the theory we do not expect the conformal part 
of the geometry to play a dynamical role. From this point of view -- in
addition to any Teichm\"uller parameters --
at most the constant mode of the conformal factor (equivalently,
the two-dimensional total area) of the spatial geometry should appear
in the Hamiltonian.

We know that at the discretized level there are transitions between
any pair of two-geometries of the same topology, that is, all
matrix elements of $\hT$ are non-vanishing. It would be 
very interesting to understand in detail how the matrix elements lose 
their sensitivity to anything 
but the Teichm\"{u}ller parameters and the total area in the continuum limit.
Although the $ABAB$-matrix model cannot be used to address the
issue of how the dependence of
the transfer matrix on the conformal factor drops out, 
solving its {\it asymmetric} version (with $\alpha_1 \not=
\alpha_2$) would determine the 
dependence of the transfer matrix (and thus the quantum
Hamiltonian) on the area of the spatial boundaries. We hope to return to
this issue in the near future.
\vspace{.4cm}

\noindent {\it Acknowledgment}.
All authors acknowledge support by the
EU network on ``Discrete Random Geometry'', grant HPRN-CT-1999-00161. 
In addition, J.A. and J.J. were supported by ``MaPhySto'', 
the Center of Mathematical Physics 
and Stochastics, financed by the 
National Danish Research Foundation.

\end{document}